# Comparing Data-Driven and Numerical Weather Prediction Forecasts of Extreme Weather: A Review

Gabriele Messori[1,2,3], Hiu-Wang Lau[1], Chenyu Dong[4], Gianmarco Mengaldo[4], Leonardo Olivetti[1,2]

1. Department of Earth Sciences, Uppsala University, Uppsala, Sweden
2. Swedish Centre for Impacts of Climate Extremes, Uppsala University, Uppsala, Sweden
3. Department of Meteorology, Stockholm University, Stockholm, Sweden
4. College of Design and Engineering, National University of Singapore, Singapore

Correspondence to: Gabriele Messori (gabriele.messori@geo.uu.se)

## Abstract

Data-driven weather forecasting models, chiefly trained on reanalysis data, now match or beat physics-based numerical weather prediction (NWP) models on most headline scores, but whether this advantage extends to extreme weather is contested. We review peer-reviewed studies that directly compare data-driven forecasts with NWP for extremes, from nowcasting to subseasonal lead times. We discuss grey literature in the outlook but do not include it in the formal review. There is good agreement on the fact that data-driven models mostly outperform NWP on temperature extremes, tropical cyclone tracks and the location of synoptic features across time scales. The evidence is mixed for extreme precipitation and feature-based extreme event detection such as extremes related to atmospheric rivers, while it leans toward better NWP performance for estimating event intensity. For record-breaking events that lie outside the training distribution, the evidence is narrower. Nonetheless, both a systematic global evaluation and several case studies find that physics-based models retain the advantage, with data-driven models underestimating both the frequency and the magnitude of records. Hybrid systems that post-process NWP with machine-learning (ML), or couple data-driven forecasts with physics-based models, deliver some of the largest and most consistent gains. We argue that headline scores, verification choices and case selection explain much of the apparent disagreement in the literature, and provide recommendations for a more trustworthy and interpretable evaluation of data-driven models for extreme and record-breaking weather.

## 1. Introduction

For over half a century, weather forecasts have been produced by numerically integrating the equations of atmospheric motion. Improvements in observations, data assimilation, model physics and computing have added roughly a day of useful lead time per decade (Bauer et al., 2015). Since 2022 this trajectory has been disrupted by data-driven models trained on decades of reanalysis, most prominently ECMWF’s ERA5 (Hersbach et al., 2020). FourCastNet (Kurth et al., 2023), GraphCast (Lam et al., 2023), Pangu-Weather (Bi et al., 2023), FuXi (Chen et al., 2023), Aurora (Bodnar et al., 2025), GenCast (Price et al., 2025) and ECMWF's own Artificial Intelligence Forecasting System (AIFS, Lang et al., 2026) are some examples of data-driven systems that both drastically reduce computational costs relative to numerical weather prediction (NWP) and outperform state-of-the-art NWP forecasts on various headline scores. For example, GraphCast outperformed the ECMWF integrated forecasting system high-resolution deterministic forecast (IFS HRES) on 90% of 1,380 verification targets (Lam et al., 2023). Similarly, the diffusion-based ensemble GenCast outperformed the ECMWF ensemble (ENS) on 97.2% of 1,320 targets (Price et al., 2025).

These headline results rest largely on standardized performance metrics averaged over the globe and over all days available for testing. However, not all weather forecasts are equally valuable for their users. In most cases, accurate forecasts of extreme events are those that come with the greatest benefit. At the same time, forecast skill for extremes is often lower than for average conditions. For NWP, the challenge comes from the fact that the physical processes driving extremes may not be the same as those associated with average weather, and that parametrisations that work well for most conditions may have degraded performance for extremes. For data-driven models, since extremes sit in the tails of the training distribution, optimization on mean-squared error or similar mean-focussed metrics rewards for optimizing overall skill at the cost of poor performance on rare events (Olivetti and Messori, 2024a). While there is broad agreement that data-driven models now outperform state-of-the-art NWP on overall skill metrics, the evidence for extreme events is less clear-cut.

An additional challenge comes from the non-stationarity of the climate system. The severity of many categories of extreme events is increasing in a warming climate, and records are being beaten much more frequently than would be expected in a stationary system (Fischer et al., 2025). Thus, a data-driven model that has learnt only from the past must be able to extrapolate. Physics-based models face their own limits, including parameterized convection, finite resolution and systematic biases, but they are not bound by the range of historical data. Whether data-driven models can extrapolate to forecast events that lie outside their training distribution is therefore a cogent question, and one on which there is still comparatively little evidence.

Here, we conduct a systematic review of studies that explicitly compare data-driven forecasts with physics-based numerical forecasts. We address two questions. The first is how well data-driven models forecast extremes that are rare but sampled in the training data. The second is how well they forecast record-breaking events, whose magnitude exceeds anything in the training period and which therefore test extrapolation. We first discuss how comparisons are made in the literature and then review evidence on each question in turn. Thereafter, we take a closer look at the performance of hybrid forecasting systems, and conclude by reconciling seemingly contrasting results and outlining future priorities for evaluation.

## 2. Systematic literature search

We searched Web of Science on the 30th October 2026 using the following search terms: “Weather forecast*" and "machine learning" and "record*"; "Weather forecast*" and "data-driven" and "record*"; "Weather forecast*" and "AI" and "record*"; "Weather forecast*" and "machine learning" and "extreme*"; "Weather forecast*" and "data-driven" and " extreme *"; and "Weather forecast*" and "AI" and " extreme*". This returned 806 records in total, and 646 after deduplication. We then selected papers by screening title and abstract based on the following requirements, all of which need to be fulfilled for inclusion: Quantitative comparison between data-driven and NWP forecasts; comparison includes performance on extreme events or on feature-based extreme event detection; comparison is performed at meteorological or subseasonal timescales (lead times up to a few weeks); comparison includes meteorological variables directly related to (near-)surface extremes. Following this screening, 75 papers were retained. We additionally applied a discretionary step for papers which included minimal analysis on extremes, applying our expert judgement to decide whether they provided useful additional information for the purposes of this review, whereby a further 2 papers were removed. 73 papers were thus retained as presenting results of direct relevance for comparison of data-driven and NWP forecasts of weather extremes. A considerable number are partly or wholly dedicated to evaluating hybrid approaches, in which ML post-processes, downscales or steers an NWP model, against the NWP forecast alone. The 73 papers range from nowcasting to seasonal prediction (although all those that we retain dealing with seasonal forecasts also present results for the shorter timescales that we consider in this review), and from global benchmarks to single events. We did not include preprints and other grey literature sources in our bibliographic search. We did however integrate the list of the 73 papers with additional references to provide context for the review and discussion of results (Fig. 1).

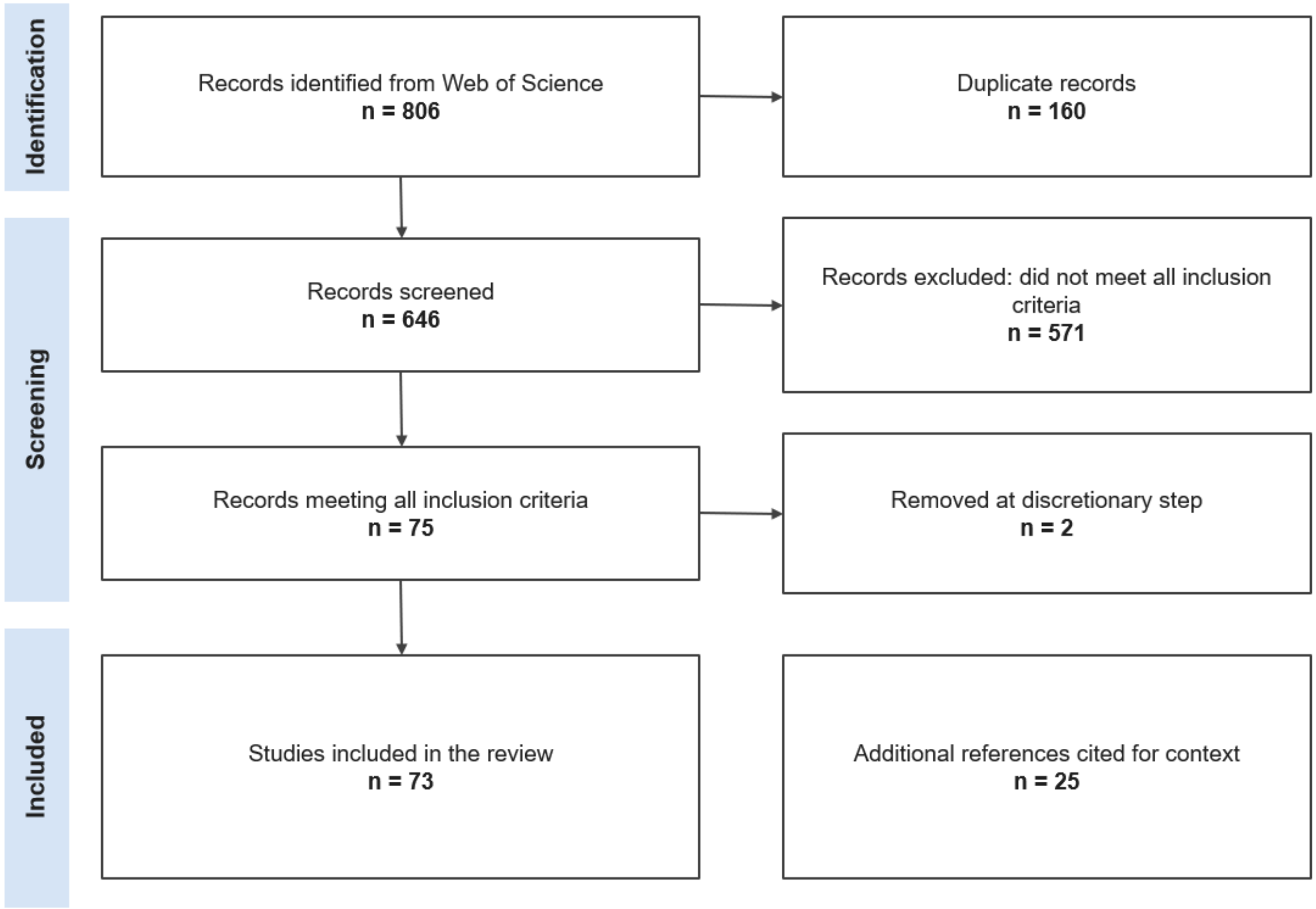

***Figure 1** Flow diagram illustrating the selection of papers included in the review.*

## 3. Considerations and challenges in comparing data-driven and numerical forecasts of extremes

Comparing data-driven and numerical forecasts of extreme events involves several important methodological considerations and challenges, and conclusions can be sensitive to the choices made in the evaluation process. We underscore the importance of accounting for the characteristics of the data-driven models, the data used for initialization and verification, and the metrics used to assess forecast skill.

Data-driven weather forecasting models differ in their deterministic or probabilistic formulation, spatial coverage and resolution, forecast lead times, and degree of specialization. A key distinction is between deterministic and probabilistic formulations. Deterministic models produce a single forecast and are often trained by minimizing mean-squared error or mean-absolute error (Kurth et al., 2023; Bi et al., 2023; Lam et al., 2023; Chen et al., 2023; Bodnar et al., 2025), which can lead to smoothing at longer lead times (Bonavita, 2024). Probabilistic models, in contrast, aim to represent a distribution of possible outcomes, allowing them to better capture forecast uncertainty and variability (Price et al., 2025; Couairon et al., 2026; Lang et al., 2026; Nai et al., 2025). Another important distinction is between global models and regional models. The latter often operate at higher spatial resolutions, potentially offering advantages in capturing the intensity and spatial structure of localized extremes (Baño-Medina et al., 2025; Nipen et al., 2026). Data-driven models also span a wide range of forecast lead times, from nowcasting to seasonal prediction (e.g. Chen et al., 2025; McAdam et al., 2025), although in this review we limit ourselves to forecasts up to subseasonal lead times. Over these timescales, the predictability of extreme events and the relevant measures of forecast skill can differ substantially. Moreover, some data-driven approaches are designed for specific forecasting tasks, such as random forests for subseasonal heatwaves (Weirich-Benet et al., 2023) or deep learning models for precipitation nowcasting (Chen et al., 2025). This diversity of data-driven models, together with the range of numerical forecasting systems used as benchmarks, adds complexity to comparing extreme weather forecasts. Depending on the forecasting task, studies may compare data-driven models against global or regional and deterministic or ensemble NWP forecasts, on timescales from hours to several weeks.

Another challenge lies in the choice of initial conditions and verification (ground truth) data. Early evaluations initialized data-driven models from ERA5 and NWP from operational analyses, which is not a like-for-like test. Operational-style comparisons that initialize both from the same analysis provide a more consistent protocol (Ben Bouallègue et al., 2024; Olivetti and Messori, 2024b). The difficulty of ensuring a "fair" comparison extends to the choice of verifying ground truth. Most global studies verify against ERA5 or the operational analysis used to initialize IFS HRES. ERA5 is the training target of the data-driven models, which can favour them, while the operational analysis may favour IFS HRES. One approach to address the mismatch between reanalysis-based training and operational forecasting is to fine-tune data-driven models on operational analyses, as has been done for GraphCast (Lam et al., 2023) and AIFS (Lang et al., 2026). Independent station observations offer an alternative verification reference (Bremnes et al., 2024), although their spatial coverage is uneven, limiting their representativeness. The importance of the verification reference is illustrated by the fact that conclusions about specific extreme event predictions, such as tropical cyclone intensity, may reverse depending on whether ERA5 or best-track data are used (Zhong et al., 2024). Similarly, data-driven models that perform well when verified against ERA5 may show no advantage over NWP models when compared against station data (Dahl et al., 2026). Related to this, case selection can also influence the conclusions. Case studies enable detailed forecast diagnostics and error analyses, but evaluating forecasts only for events selected by their observed outcomes can introduce selection bias and favour over-forecasting, referred to as the forecaster's dilemma (Lerch et al., 2017; Pasche et al., 2025).

The final element that we highlight is the choice of scoring metric, especially for deterministic forecasts. RMSE penalizes a sharp forecast that misplaces a feature twice, once where it is forecast and once where it occurs. This double penalty rewards the smoothing typical of deterministic data-driven models (Bonavita, 2024; Bonavita and Geer, 2026). Several approaches have been proposed to mitigate this issue. Gneiting et al. (2026) argue for the use of a continuous ranked probability score for deterministic forecasts that subjects forecast model output to the same statistical post-processing before comparison. Brenowitz et al. (2025) propose an evaluation of probabilistic forecasts based on ensembles built with a lagged-initialization technique, which can be applied to deterministic forecasts. Other evaluations have used scores focusing on tail values or on binary extreme – non-extreme forecasts, such as Q–Q plots, the threshold weighted Continuous Ranked Probability Score (twCRPS, Gneiting and Ranjan, 2011), the critical success index (CSI) or the symmetric extremal dependence index (SEDI).

## 4. Performance of stand-alone data-driven forecasting models

An important distinction for evaluating data-driven weather forecasting models, is between extremes within the range of the training data and record-breaking events that require extrapolation beyond it. We review the evidence for these two categories of events separately in the following subsections.

### 4.1 Within-sample extremes

Across the literature corpus, data-driven models are competitive with NWP for extremes sampled in the local climatology, but their advantage is smaller than on average scores and depends strongly on variable, region, lead time and metric. Table 1 summarizes where the available evidence converges and where it conflicts.

**Temperature and wind.** The most systematic evidence concerns near-surface temperature and wind extremes. For deterministic forecasts, GraphCast showed higher precision and recall than IFS HRES for extreme heat and cold at most lead times (Lam et al., 2023). Moreover, an operational-like test of Pangu-Weather against IFS HRES found comparable skill for European winter cold extremes and significantly better skill for Pangu-Weather in forecasting summer temperatures between the 75th and 90th percentiles (Ben Bouallègue et al., 2024). In a semi-operational comparison, Pangu-Weather and GraphCast were comparable on tail performance to IFS HRES overall, showing lower error at shorter lead times but becoming comparatively worse beyond five to seven days (Olivetti and Messori, 2024b). The same study found a pronounced meridional pattern: data-driven models outperformed IFS HRES on temperature extremes in the tropics but were poorer at higher latitudes and on the eastern sides of ocean basins. The latitudinal dependence may be related to the use of latitude-based weights, which promote the best performance around the Equator. Olivetti and Messori (2025) found similar results using AIFS: at many locations AIFS outperformed IFS HRES for temperature and wind extremes, yet the latter displayed lower error at many land points, especially at longer lead times.

Regionally, data-driven models initialized from Korean analyses beat the operational Korean forecast model for temperature extremes (Oh et al., 2026), and data-driven heat-stress forecasts outperformed GFS in southern South America for lead time of up to three days (Collazo et al., 2026). For the Southwest Europe 2023 heatwave Aurora showed comparable performance to IFS up to roughly one week, while for the February 2021 Texas freeze, it matched GFS at a 1-day lead time but displayed increasing underperformance by lead time 7 days (Huang et al., 2026b). Pasche et al. (2025) instead find for the same North American winter storm that, for several lead times, Pangu-Weather and GraphCast outperform HRES. For wind, the stretched-grid Bris model beat IFS in equitable threat score for strong winds over Norway, although it underperformed MEPS for the strongest windspeeds (Nipen et al., 2026). For the extratropical windstorm Poly, however, Bris missed the strongest winds near the cyclone centre and violated key dynamical balances, thus comparing unfavourably with MEPS (Pasquini et al., 2026). Over the South China Sea, Pangu-Weather had lower wind RMSE than ECMWF but systematically underestimated boundary-layer jet intensity and extent, whereas ECMWF overestimated them (Li and Du, 2026). Over California, data-driven models were comparable to GFS and IFS HRES against station data for 10m winds, but clearly underperformed the high-resolution West-WRF (Dahl et al., 2026). In a cost–loss analysis of the economic value of AIFS and IFS HRES extreme event forecasts of temperature and windspeed, the preferred model depended on the location, the forecasted variable and the ratio of protection cost to loss (Olivetti et al., 2026). IFS HRES's higher sensitivity made it more valuable where protection is cheap relative to losses, while AIFS's higher specificity made it more valuable where protection is costly.

Beyond two weeks, data-driven models increasingly compete with numerical systems, but the forecasting of extremes remains generally weak. Random forests forecasting central European heatwaves compete with the ECMWF’s numerical model hindcast ensemble mean beyond two weeks, yet they captured extreme values poorly (Weirich-Benet et al., 2023). For the western

United States, data-driven models beat SubX numerical models on average but made most of their worst forecasts during cold waves, which the numerical models predicted reasonably well (He et al., 2022). FuXi-S2S gave skillful Eurasian cold-event forecasts 3–4 weeks ahead, longer than numerical systems, but was less skillful for an atypical circulation with few training analogues (Zhu and Dai, 2026).

For ensemble forecasts, the evidence is more limited. GenCast had higher Brier skill than ENS for extreme heat and cold, and high winds (Price et al., 2025). Very large ensembles from spherical Fourier neural operators could reproduce ECMWF's extreme forecast index with similar reliability and discrimination, while lagging ENS by 12–18 h on overall scores (Mahesh et al., 2025).

**Precipitation.** The number of studies evaluating the performance of data-driven models for precipitation is smaller than for temperature and winds. Many studies have a regional focus and the evaluations are largely limited to deterministic forecasts. AIFS had overall lower 1–5-day precipitation RMSE than IFS over China; the best-performing model for precipitation extremes depended on the chosen evaluation metric (Pan et al., 2025). Over Bangladesh, GraphCast had the highest correlation and lowest RMSE for 1–3-day rainfall but lower CSI and higher false-alarm ratios than ECMWF and GFS for extreme events (Hasan et al., 2026). For Taiwan, a nowcasting data-driven model showed improved performance relative to the Rapid Weather Research and Forecasting numerical model for heavy precipitation case studies (Chen et al., 2025). Finally, data-driven approaches have also shown consistent improvements over numerical models for prediction of winter precipitation type, such as freezing rain, in China (Lang et al., 2023; Wen et al., 2024).

Approaches that explicitly target fine scales provide a similar picture. A latent diffusion model with a 4 km resolution driven by Pangu-Weather forecasts, outperformed GFS for several extreme precipitation case studies (Li et al., 2025a). While the diffusion model provides an ensemble, no probabilistic comparison to the numerical model was performed. A 6-km regional data-driven model captured atmospheric-river precipitation extremes better than both IFS and its own 31-km global parent model, and performed competitively with the regional West-WRF (Baño-Medina et al., 2025). FuXi-Extreme, a diffusion model that enhances fine-scale details in FuXi surface fields, outperformed both FuXi and IFS HRES on CSI and SEDI for extreme precipitation, wind and temperature in global evaluations (Zhong et al., 2024). Super-resolution downscaling of FuXi precipitation to 1 km outperformed IFS HRES over China, although all models, IFS HRES included, underestimated the most intense rainfall (Li et al., 2025b).

**Feature-based extreme event detection.** We also review the performance of data-driven and NWP forecasts for circulation features associated with extremes. Atmospheric rivers (ARs) show a clear distinction between grid-point and phenomenon-based performance. GraphCast improved integrated vapour transport (IVT) skill over IFS HRES (Lam et al., 2023), and data-driven models outperformed NWP on grid-point IVT metrics and AR location (Swenson et al., 2026). Object-based evaluations, however, showed that data-driven models identify ARs which are too small, too weak and too smooth (Swenson et al., 2026). When using geometric detection algorithms, IFS HRES had a clear advantage in AR detection, especially in the mid-latitudes and at short lead times, despite data-driven models displaying comparable or better IVT RMSE (Lopez-Marti et al., 2025). Along the US West Coast, IFS HRES had higher AR detection skill than five data-driven models during the first four forecast days. Aurora displayed strong variable-specific RMSE, but was worst at detecting ARs (Davis et al., 2026).

For extratropical systems, data-driven models outperformed traditional NWP by up to 20–30% in RMSE for high-impact US East Coast winter storms. They also captured conveyor-belt structure 7–10 days ahead (Spallone and Greybush, 2026). Similarly, they matched IFS HRES on synoptic evolution and frontal structure for two European storms (Jiménez-Esteve et al., 2026). GraphCast and Pangu-Weather matched or exceeded IFS HRES RMSE for convective instability and shear forecasts, but did not consistently beat it on fractional skill score (Feldmann et al., 2024). The authors also underscore that individual case studies can easily lead to conclusions that differ from those drawn from a larger sample of events.

Data-driven models perform consistently well on forecasting tropical cyclone tracks. GraphCast, GenCast, FuXi and Aurora generally beat IFS HRES, ENS or official forecasts on tracks (Lam et al., 2023; Price et al., 2025; Zhong et al., 2024; Bodnar et al., 2025). Similarly, physically constrained perturbations of FuXi produced ensemble track forecasts that beat ENS (Pu et al., 2025). Cyclone intensity is however an area of weakness. FuXi and FuXi-Extreme underperformed IFS HRES on cyclone intensity, verified against best-track data (Zhong et al., 2024). For Hurricane Helene (2024), three data-driven models matched GFS on track forecasts, but underestimated low-level winds and the cyclone's central pressure, whereas GFS did not (Sudler et al., 2026). For Typhoon Bebinca, a data-driven model reduced track error relative to WRF but performed worse on intensity (Huang et al., 2026a). Data-driven models over East Asia likewise underestimated rainfall associated with typhoons (Oh et al., 2026).

We conclude by briefly considering forecasts of circulations associated with surface extremes. Two data-driven models outperformed ECMWF's numerical model for forecasting persistent stratospheric perturbations at lead times beyond about 25 days, yet the numerical model offered better performance at shorter lead times (de Fondeville et al., 2023). Similarly, Peings et al. (2026) found that two data-driven weather models matched ECMWF's numerical model for predicting the Madden–Julian Oscillation at lead times of three weeks and beyond.

*Table 1. Summary of agreements and disagreements in comparisons of data-driven and numerical forecasts of within-sample extremes, including example references. Articles in bold include comparisons between ensemble-based data-driven and numerical models.*

| Phenomenon | Data-driven models usually better | NWP models usually better | Example sources of disagreements |
|---|---|---|---|
| Temperature extremes | Tropics; summer heat; short to medium range; relative economic value (Lam et al., 2023; Olivetti and Messori, 2024b; Ben Bouallègue et al., 2024; **Price et al., 2025**) | High latitudes; many land points at longer leads (Olivetti and Messori, 2024b; Olivetti and Messori, 2025) | Region, lead-time, evaluation score |
| Wind extremes | Grid-point based scores in some regions; relative economic value (Olivetti and Messori, 2024b; **Price et al., 2025**; Nipen et al., 2026) | Peak winds in storms; jet intensity (Pasquini et al., 2026; Li and Du, 2026) | Smoothing of small-scale maxima |
| Heavy precipitation | High-resolution forecasts; location of maxima (Zhong et al., 2024; Li et al., 2025a; Baño-Medina et al., 2025) | Deterministic forecasts at extreme thresholds (Hasan et al., 2026) | Model resolution; observational reference |
| Atmospheric rivers | IVT RMSE; AR location (Lam et al., 2023; Lopez-Marti et al., 2025; Swenson et al., 2026) | Object-based detection; AR intensity and size (Lopez-Marti et al., 2025; Davis et al., 2026; Swenson et al., 2026) | Grid-point versus feature-based verification |
| Tropical cyclones | Track (Lam et al., 2023; **Price et al., 2025; Pu et al., 2025**) | Intensity (Zhong et al., 2024; Sudler et al., 2026; Huang et al., 2026a) | ERA5 versus best-track reference |
| Subseasonal lead times | Longer lead times (**Weirich-Benet et al., 2023**; Zhu and Dai, 2026) | Shorter lead times; atypical events and cold waves (**He et al., 2022; Zhu and Dai, 2026**) | Lead times, level of rarity of the events |

## 4.2 Record-breaking events

Record-breaking events pose a different question from within-sample extremes, as they require data-driven models to extrapolate beyond the conditions represented in their training data. A 95$^{th}$ or 99$^{th}$-percentile heatwave may have many analogues in a training sample consisting of several decades of reanalysis, yet a record-breaking one does not. Forecasting it requires the model to produce values outside the range it has been trained on, and there is no *a priori* reason to expect that an emulator fitted to historical data should be skillful at this. This was illustrated in the early days of data-driven forecasts by Scher and Messori (2019), who showed that a neural network which can reproduce almost perfectly the Lorenz-63 attractor when seeing the full attractor in its training set, performs very poorly on any part of the attractor that is removed from the set. More recently, Sun et al. (2025) conducted an analogous investigation in a more realistic weather forecasting setting. The authors showed that the FourCastNet model trained without strong tropical cyclones struggles to predict unseen Category 5 storms. Based on our literature search and selection criteria, only a handful of the 73 studies address record-breaking events explicitly. The available evidence points toward a generally superior performance of NWP models.

Zhang et al. (2026) provide the most comprehensive test currently available of out-of-sample performance. They compiled grid-point exceedances of the training-period record in the test years 2018 and 2020. For 2020 alone this gave 162,751 heat, 32,991 cold and 53,345 wind records over land. Data-driven models mostly outperformed IFS HRES when considering all forecast cases, consistent with earlier benchmarks, but the ranking was reversed when restricting the evaluation to record-breaking events. IFS HRES outperformed GraphCast, Pangu-Weather and FuXi, and the former models' operational versions, for hot, cold and wind records at nearly all lead times. The data-driven models generally underestimated both the frequency and the intensity of records, with errors growing for events that exceeded historical records by larger margins. The authors attribute this to a limited ability to extrapolate beyond the training sample, a limitation that becomes increasingly important as a changing climate favours the occurrence of new records (Fischer et al., 2025). The authors recognize the challenges in choosing a fair verifying ground truth. In addition to reporting results for ERA5-trained data-driven models against ERA5 reanalysis, and HRES against its own lead time 0 analysis, they also report the operational versions of GraphCast and Pangu-Weather against the HRES analysis.

Case studies of individual record-breaking events tell a comparable but more nuanced story. For the 2021 Pacific Northwest heatwave, which broke national records by several degrees, GraphCast and Pangu-Weather achieved accuracy similar to IFS HRES locally, but underperformed when errors were aggregated over space and time (Pasche et al., 2025). At short lead times IFS HRES displayed smaller errors, unlike for ordinary summers, which the authors interpret as an example of poor data-driven extrapolation. At the same time, IFS HRES underperformed the data-driven models in forecasting the sharp temperature drop after the heatwave peak. For the 2023 South Asian humid heatwave, the data-driven models underestimated the highest heat-danger levels over Bangladesh (Pasche et al., 2025). However, the comparison was hindered by the fact that none of them predicted surface humidity directly. For the exceptional June 2023 heatwave in northern China, a generative data-driven ensemble (GenEPS) achieved better probabilistic forecast skill than ECMWF-ENS, although both ensemble means underestimated the heatwave's intensity in the northeast (Nai et al., 2025).

In some cases, data-driven models may capture the timing and location of record-breaking events without reproducing their physical characteristics. The December 2022 cold front over the US High Plains produced record temperature falls of 13.8 °C in 10 seconds and more than 34 °C in 12 h (Schumacher et al., 2025). Three to four days ahead, data-driven models predicted the timing and location of the front better than operational physics-based models, which forecasted an excessively slow frontal propagation. The data-driven models did not, however, represent the extreme gradients. High-resolution NWP showed sharp gradients, albeit still underpredicting the observed intensity, and a 600-m simulation improved the structure while remaining too slow. Schumacher et al. (2025) speculate that the data-driven models' better timing reflects the typical, faster frontal speed learned from historical data. If so, this reliance on historical patterns may also prevent them from reproducing what was atypical about the event. The July 2021 Henan rainfall in China, which broke the national hourly precipitation record, shows a similar difference between data-driven and NWP models (Li and Zhao, 2026). GraphCast and FuXi captured the large-scale drivers of the event, including the enhanced western North Pacific subtropical high, Typhoon In-Fa and moisture convergence. GraphCast was the better-performing of the two data-driven models, and outperformed IFS HRES both in describing the temporal evolution of the precipitation when initialized a few days prior to the event onset, and in identifying its spatial location. However, GraphCast underestimated the accumulated precipitation during the event even for forecasts initiated at the event onset, underperforming IFS HRES in this respect.

A recent study suggests that AI weather models can leverage knowledge learned from other regions to predict locally unprecedented extreme events. Sun et al. (2026) note that GraphCast and AIFS forecast the 2024 Dubai rainfall, roughly twice the regional training maximum, up to eight days ahead. They attribute this to dynamically similar events in other regions of the training set (Sun et al. 2025; Sun et al. 2026). This highlights the complexity hidden in the label of “record-breaking”, as something that is unprecedented at a location is not necessarily unprecedented in the overall training sample of a global model. Generalization across regions – which the authors term “translocation” – may succeed when good analogues for a locally record-breaking event exist elsewhere.

## 5. Performance of hybrid forecasting approaches

The complementary strengths of data-driven and numerical forecasts motivate hybrid approaches that combine the two forecasting paradigms. Unlike standalone data-driven forecasting models, these approaches retain a physics-based component. Hybrid forecasting has a longer history than standalone data-driven weather forecasting, notably through machine-learning post-processing of NWP output. Across these studies, the hybrid approaches show systematic improvements over standalone NWP forecasts.

The largest group of studies uses data-driven approaches to post-process, calibrate or downscale NWP output for a target variable, and compares the result with the raw NWP model. For heavy precipitation, random forests trained on reforecasts added significant skill over raw GEFS and ECMWF ensembles for locally extreme rainfall across the contiguous United States (Herman and Schumacher, 2018). Similar gains were found for calibrated short-range ensembles (Loken et al., 2019), ecPoint-style gradient-boosted calibration in Australia (Trotta et al., 2024) and a hybrid dynamical-ML model which combines forecast atmospheric predictors for precipitation extremes in Italy (Grazzini et al., 2024). U-Net post-processing reduced the RMSE of a high-resolution WRF configuration by 13–16%, adding more than a day of predictive skill and performing best for the prediction of extreme events (Badrinath et al., 2023). A graph neural network improved ECMWF rainfall over East China at all intensities (Peng et al., 2023). A conditional generative adversarial network improved IFS HRES rainfall forecasts over East Africa, where convection is poorly simulated, by postprocessing the deterministic forecasts into an ensemble (Antonio et al., 2025). A similar concept was used by Chapman et al. (2022) and Higgins et al. (2026) to improve IVT forecasts associated with ARs over the North Pacific and Western North America, building on West-WRF deterministic forecasts. The results were comparable to or outperformed GEFS and ECMWF ensemble forecasts. This followed previous work showing improvement of deterministic IVT forecasts by convolutional neural network postprocessing of GFS forecasts, with RMSE predictability gains of 1-2 days (Chapman et al., 2019). Ghazvinian et al. (2024) instead postprocessed West-WRF ensemble forecasts using deep-learning, achieving improved reliability and skill of heavy-to-extreme precipitation forecasts in the Western United States.

Comparable gains appear for other hazards. For wind gusts in Northeast US storms, gradient boosting on WRF output beat WRF's own post-processor (Jahan et al., 2024). An evidential neural network reduced gust RMSE by 47% relative to WRF and supplied calibrated uncertainty (Jahan et al., 2025). Ensemble learning on ECMWF IFS HRES reduced temperature and wind-speed errors by 48% and 29% over complex terrain during the Beijing 2022 Winter Olympics (Sun et al., 2024). ML combined with WRF variables improved snowfall accumulation over the US Northeast (Khaira et al., 2024). A follow-up study found that WRF combined with gradient boosting outperformed WRF’s own snowfall diagnostic, with higher temporal resolution being key to reduce error during rapidly intensifying storms, although model ranking partly depended on the choice of evaluation metric for heavy snowfall (Khaira and Astitha, 2025). ML post-processing similarly improved tropical cyclone intensity in Indian and Chinese ensemble systems (Kumar et al., 2023; Feng et al., 2024). It also improved maximum and minimum temperature extremes over India (Singh et al., 2026), visibility in Korea (Kim et al., 2022) and road surface temperatures near freezing in Finland (Karsisto et al., 2026). At subseasonal ranges, probabilistic bias correction applied to both the ECMWF dynamical model and AIFS improved performance for the vast majority of forecasts, including those for extremes. The hybrid system performed best for all variables and lead times in ECMWF's 2025 real-time competition, ahead of six operational centres and AIFS itself (Guan et al., 2026). Pan et al. (2026) found similar results for week 3–4 deterministic and probabilistic extreme-heat forecasts over the Yangtze River Basin, using a U-Net based model building on numerical forecast input.

A second group of studies couples data-driven and physics-based models directly. Spectrally nudging the large scales of the Canadian GEM model toward GraphCast improved GEM's skill, while retaining fine-scale detail and a full set of physically consistent variables (Husain et al., 2025). The same approach applied to CMA-GFS and FuXi successfully combined CMA-GFS's skill in forecasting rainfall intensity and cyclone intensity and FuXi's ability to predict the large-scale circulation, precipitation patterns and cyclone tracks (Su et al., 2026). Useful threat scores for 400-mm rainfall during the July 2023 North China floods were extended from 3 days to 8.5 days when WRF was forced with Pangu-Weather boundary conditions rather than from GFS (Xu et al., 2024). Similarly, WRF forced by XGBoost-extended Pangu-Weather reduced both temperature and wind RMSE relative to IFS HRES (Jin et al., 2025), although no systematic evaluation for extremes was performed. For Typhoon Bebinca, weighting numerical and data-driven forecasts in real time improved the cyclone track over the numerical forecast while performing comparably for intensity (Huang et al., 2026a). Consistent with this, combining IFS HRES and Pangu-Weather boundary-layer jets exploited the forecasts' opposite biases (Li and Du, 2026).

Additional studies that were identified in our bibliographic search present results consistent with the above. These covered extreme rainfall and temperature extremes on multiple timescales (Shastri et al., 2017; Liu et al., 2024; Grönquist et al., 2021; Landry et al., 2024; Yao et al., 2025; Liu et al., 2026). The generally strong performance of the hybrid forecasts however does not mean that skill cannot degrade under extreme or atypical conditions.

Finally, there are studies embedding data-driven modules within physics-based models, typically as emulators of expensive parameterizations. For example, a neural-network radiation emulator in the Korean operational model ran up to >2,000 times faster than the control radiation scheme without significantly degrading precipitation or temperature forecasts (Song and Roh, 2021). However, the targeted application of such approaches to extreme event forecasting is limited. These studies nonetheless show a route by which ML can help make higher-resolution NWP forecasts computationally manageable.

## 6. Reconciling contrasting results and outlook

The evidence reviewed above reveals a complex picture of the relative strengths and limitations of data-driven and numerical forecasts of extreme weather (Fig. 2). We argue that the apparent contradictions in the literature comparing data-driven and NWP forecasts of extremes can be largely explained by methodological and data choices. Recognizing these differences is essential for interpreting and reconciling the available evidence.

The first distinction is between within-sample extremes and out-of-sample records. Data-driven models are generally competitive for within-sample extremes. However, they show larger limitations for out-of-sample events, for which physics-based models are still the strongest performers (Zhang et al., 2026). Some of the contrasting results in the literature may therefore be explained by differences in the rarity of the events being considered.

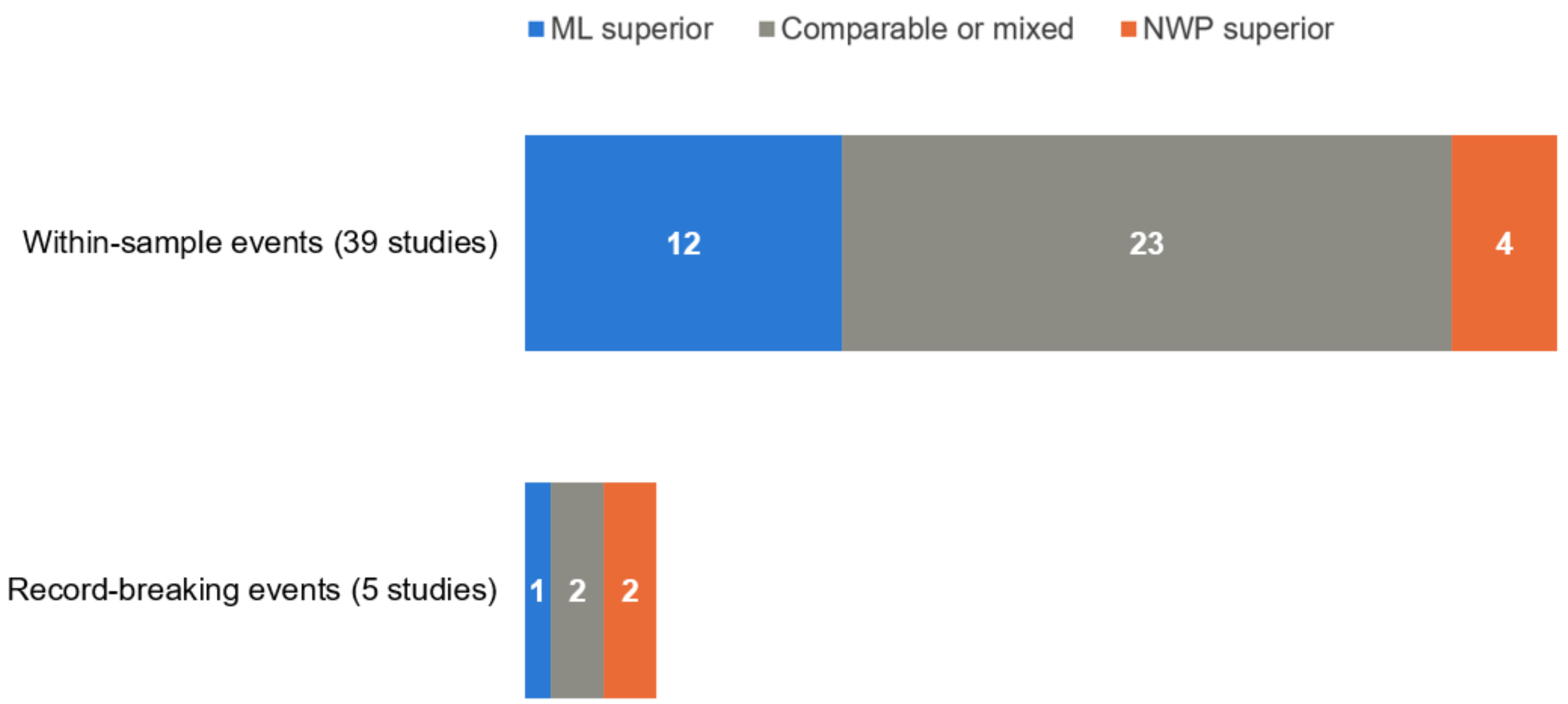

***Figure 2*** *Classification of 43 papers considered in this review, excluding hybrid approaches. Some papers are included in both the within-sample and record-breaking categories. "Comparable or mixed" covers studies reporting similar skill and those in which the best-performing model heavily depends on variable, metric or lead time*.

The second source of discrepancies is how data-driven and numerical forecasts of extremes are compared. This includes the choice of scoring metric, the models that are considered in the comparison and the evaluation set-up (verifying ground truth, model initialization etc.). Grid-point RMSE, anomaly correlation and some threshold-based scores reward correct location and timing, for which data-driven models perform strongly. NWP models appear to still have an edge for predicting peak magnitude, sharp gradients, feature size and for ensuring physical consistency. As example of these different strengths, the same data-driven models that outperform NWP on IVT RMSE can perform more poorly on AR detection (Lopez-Marti et al., 2025; Davis et al., 2026; Swenson et al., 2026). However, many of the object-detection techniques used in the comparisons were originally designed for the output of numerical models and reanalysis data, which may introduce a bias in favour of NWP models. The choice of models in the comparison also has major implications. Deterministic data-driven models trained with mean-squared error increasingly smooth extremes with lead time (Bonavita 2024). Generative, diffusion-based, ensemble and high-resolution regional models retain fine-scale detail, and often reverse model ranking for precipitation and wind forecasts (Zhong et al., 2024; Price et al., 2025; Li et al., 2025a; Baño-Medina et al., 2025). Additionally, evaluation set-up can also play a major role, with results sometimes reversing depending on whether ERA5, best-track or station data are chosen as verifying ground truth (Zhong et al., 2024; Dahl et al., 2026). Finally, results vary widely with latitude, across regions (Olivetti and Messori, 2024b) and between single case studies and longer-term statistics (Feldmann et al., 2024).

These factors provide guidance for future work. We recommend that evaluations of extreme event forecasts report results separately for within-sample and for out-of-sample events, or stratify the results based on level of event rarity. Comparisons should use identical initial conditions and a reference independent of all models, such as station observations, and they should account for the forecaster's dilemma (Lerch et al., 2017). This can be achieved for example by presenting standard metrics and assessments of performance across the whole distribution (e.g. Q–Q plots), metrics of physical consistency (e.g. Bonavita, 2024) or weather maps, and by utilizing proper scoring rules. Use of shared benchmarks for extremes and records, analogous to WeatherBench 2 for average skill (Rasp et al., 2024), would make different studies directly comparable. In this respect, we are aware of the as yet unpublished Extreme Weather Bench initiative, which is an important step in this direction.

Another important area of future work is continued investigation of whether generative and ensemble data-driven models, which retain tail variance, can successfully extrapolate to record-breaking events. Almost all record-breaking evaluations use deterministic models trained with mean-squared error. These models often place the event correctly, because they perform well at predicting the large-scale circulation. However, they systematically underestimate the events' magnitude. The few existing published studies on generative and ensemble data-driven models show promising results (e.g., Price et al., 2025; Nai et al., 2025; Lang et al., 2026), but those studies are often by the model developers themselves, and do not systematically investigate record-breaking events. Independent evaluations are largely lacking. We are aware of several contrasting results on out-of-sample events in the grey literature, and we look with interest as to which will be accepted in the peer-reviewed literature. A related open question is how fast the new data-driven architectures will erode the advantage that NWP still appears to hold for record-breaking events. Training on climate-model simulations of warmer climates, applying physical constraints and fine-tuning on high-resolution analyses are additional pathways to improve out-of-sample forecasts that remain largely untested.

A clear point of agreement in the literature is that the combination of physics-based models with data-driven models to post-process, calibrate or downscale their output often results in sizeable performance improvements for a range of different extremes across multiple time scales. Such hybrid approaches appear as one of the lowest-hanging fruits to improve currently operational extreme weather forecasts, as recently shown by Guan et al. (2026). Thus, systematic comparisons of these approaches, including independent evaluations across regions, event types and forecast ranges beyond those examined in the original model-development studies, would be a valuable direction for future research.

To conclude, data-driven models are already leading for forecasts of several types of within-sample extremes, notably for temperature, although their advantage over NWP is not systematic and depends on multiple factors. Data-driven models also show strong skill in forecasting the large-scale setting and timing of extremes, tropical cyclone tracks and, at subseasonal ranges, early signals of heat and cold extremes. Physics-based NWP models generally remain best performing for the magnitude of very intense events and for record-breaking events. Given the rapid evolution of the field, the authors would however not be surprised to see further gains of the data-driven models in the near future.

## Acknowledgements

The literature search was conducted manually by the authors using Web of Science. The deduplication of references and the selection for inclusion in the review was assisted by Claude, as was the drafting of the manuscript and of its display items. All AI-assisted steps occurred under human supervision and all AI outputs were manually edited by the authors, who take full responsibility for the contents of the manuscript.

GM, LO and HWL acknowledge funding by the European Union Horizon 2020 and Horizon Europe research and innovation programmes through the European Research Council (grant numbers 948309, CENÆ and 101231771, DYCLEX). GM and LO further acknowledge the Swedish Research Council Vetenskapsrådet (grant no. 2022-06599).

## Data availability statement

No new data were created for this review, and all information presented comes from the peer-reviewed literature.

## Conflict of interests

The authors have no conflicts of interest to declare.